\documentclass[a4paper,12pt]{article}
\usepackage[english]{babel} 
\usepackage[T1]{fontenc}
\usepackage{float}
\usepackage{setspace}
\usepackage{appendix}
\usepackage{parcolumns}
\usepackage{stmaryrd}
\usepackage{cite}
\usepackage{times}
\usepackage[OT1]{fontenc}
\usepackage{type1cm}
\usepackage{url}
\usepackage{subfigure}
\usepackage{braket}
\usepackage{graphicx}
\usepackage{sidecap}
\usepackage{xspace}
\usepackage{tikz}
\usepackage{pgflibraryarrows}
\usepackage{pgflibrarysnakes}
\usepackage{indentfirst}
\usepackage[mathscr]{eucal}
\usepackage{geometry}
\usepackage{booktabs}
\usepackage{indentfirst}
\usepackage{bm}
\usepackage{multirow}
\usepackage{dcolumn}
\usepackage{graphicx}
\usepackage{mathrsfs}  
\usepackage{enumerate}
\usepackage[utf8]{inputenc}
\usepackage{amsmath}
\usepackage{amssymb}
\usepackage{amsfonts}
\usepackage{amsthm}
\usepackage{mathrsfs}
\usepackage{makeidx}
\usepackage{graphicx}
\usepackage{ulem}			

\makeatletter
\@addtoreset{equation}{section}

\makeatother

\def\det{{\rm det}}

\def\Tr{{\rm Tr}}

\newcommand{\be}{\begin{equation}}
\newcommand{\ee}{\end{equation}}
\newcommand{\bea}{\begin{eqnarray}}
\newcommand{\eea}{\end{eqnarray}}
\newcommand{\vs}[1]{\vspace{#1 mm}}

\begin{document}

\vs{3}
\begin{center}
{\Large\bf Quantum fields without Wick rotation}
\vs{8}

{\large
Alessio Baldazzi\footnote{e-mail address: abaldazz@sissa.it}\ 
Roberto Percacci\footnote{e-mail address: percacci@sissa.it}\ 
and Vedran Skrinjar\footnote{e-mail address: vskrin@sissa.it}
\vs{8}
}

{International School for Advanced Studies, via Bonomea 265, I-34136 Trieste, Italy}
{and INFN, Sezione di Trieste, Italy}

\vs{5}
{\bf Abstract}
\end{center}
We discuss the calculation of one-loop effective actions
in Lorentzian spacetimes, based on a very simple application
of the method of steepest descent to the integral over the field.
We show that for static spacetimes this procedure
agrees with the analytic continuation of Euclidean calculations.
We also discuss how to calculate the effective action
by integrating a renormalization group equation.
We show that the result is independent of arbitrary choices
in the definition of the coarse-graining and we see
again that the Lorentzian and Euclidean calculations agree.
When applied to quantum gravity on static backgrounds, 
our procedure is equivalent to analytically continuing time
{\it and} the integral over the conformal factor.

\section{Introduction}

The path integral of a Lorentzian Quantum Field Theory (QFT) is
\be
Z_L(g)\equiv e^{i\Gamma_L(g)}
=\int d\phi\, e^{i S_L(\phi,g)}\ ,
\label{ZL}
\ee
where $g$ may denote external parameters or background fields.
It is often said that this expression is ill-defined
because of the oscillating character of the exponential.
This is then fixed by turning time to Euclidean time.
One defines a Euclidean action $S_E$ by
\be
iS_L\Big|_{t=-it_E}=-S_E\ .
\label{WA}
\ee
and thereby converts the functional integral (\ref{ZL}) 
to the Euclidean functional integral
\be
Z_E(g)\equiv e^{-\Gamma_E(g)}
=\int d\phi\, e^{-S_E(\phi,g)}\ .
\label{ZE}
\ee
Since the Euclidean action is typically positive definite,
this is now a better-defined object.
More precisely, the integral over each field mode is now convergent.
The functional integral still needs regularization and renormalization,
because of the infinite number of field modes.
In the end the results can be analytically continued back to
real time, which means that the Lorentzian and Euclidean
Effective Actions (EA) are related in the same way as the
classical actions:
\be
i\Gamma_L\Big|_{t=-it_E}=-\Gamma_E\ .
\label{WEA}
\ee

This procedure is much less clear
when the background metric is curved.
In a gravitational context, one would like to preserve
invariance under coordinate transformations.
Since time is now merely a coordinate,
one has to make a choice of which time 
should be analytically continued.
The problematic nature of this choice has been emphasized
by Visser \cite{Visser:2017atf}.

The procedure is even more problematic when gravity itself
is treated as a quantum field.
In a one-loop calculation, where the graviton field can be seen
as a free quantum field propagating on a curved background,
the Lagrangian of the spin-zero degree of freedom of the metric 
has opposite sign relative to the spin-two degree of freedom.
The sign of the Hilbert action is chosen so that the latter
has the correct sign (so that free gravitons in flat space
carry positive energy) and therefore the spin-zero field
has negative energy.
This in itself is not problematic, because the spin-zero field
does not propagate, but it means that while the Euclidean integral
over the spin-two degrees of freedom is exponentially damped,
the integral over the spin-zero degrees of freedom
is exponentially divergent.

This can be seen also at non-perturbative level:
the spin-zero field is related to the conformal part of the metric,
and the Hilbert action can be made arbitrarily negative
by performing a highly oscillating conformal transformation
\cite{Gibbons:1978ac}.
\footnote{Insofar as the conformal factor is a gauge degree of
freedom, one may doubt that this is significant.
In fact, the conformal factor is absent in unimodular gravity.
However, in that case another scalar
component of the graviton has wrong-sign action.
The existence of a scalar with wrong-sign action
is a gauge-independent statement.}
The standard way out, that we shall refer to as
the ``Cambridge prescription'', is to rotate the
integration over the conformal factor
in the complex plane, to make the integral over the
conformal factor convergent.

Returning to general QFT in curved spacetime,
there is an alternative to analytically continuing time,
that is analytically continuing the metric.
In an ADM decomposition, this can be thought of as
analytically continuing the lapse.
Even more generally, one can think of a one-parameter
family of metrics:
$$
g_{\mu\nu}(\sigma)=
g_{\mu\nu}+(1+\sigma)X_\mu X_\nu
$$
where $g_{\mu\nu}$ is a Lorentzian metric
and $X_\mu$ is a unit timelike vectorfield in the metric
$g_{\mu\nu}$.
For $\sigma=-1$ one has the original Lorentzian metric
while for $\sigma=1$ one has a Euclidean metric.
This procedure has been used in \cite{Candelas:1977tt}
and its advantages extolled in \cite{Visser:2017atf}.
One of the advantages is that all the metrics are
defined on the same underlying manifold.
This puts strong restrictions on the class of manifolds
that one may have to sum over,
and this in turn is known to greatly improve the
definition of the path integral \cite{Ambjorn:1998xu,cdt}.

On the other hand, this definition is also not free of
ambiguity: there is no preferred choice
of vectorfield $X_\mu$.
One may try to restrict this freedom by imposing some
additional conditions, such as 
mapping Einstein manifolds to Einstein manifolds,
or preserving the number of Killing vectors of the metric.
However, it can be seen that in some cases these requirements 
clash with the requirement of preserving the
manifold structure \cite{baldazzi}.

For this reason, in the present paper we shall explore
an alternative procedure, where no Wick rotation is performed.
By this we mean that neither time nor the metric are changed.
Instead, an analytic continuation
is performed on the quantum field itself.
More precisely, one can choose a contour in the integral
over the field such as to make the integral 
over each mode convergent.
This procedure is inspired by recent work on quantum cosmology \cite{Feldbrugge:2017kzv}, 
where the integral over the lapse is made convergent by using 
Picard-Lefschetz theory.
Here we shall discuss only the case of quadratic actions,
where it is enough to use the steepest descent definition 
of the Fresnel integrals, which is actually 
the simplest application
of Picard-Lefschetz theory and, as we shall mention in section 6,
is all that one needs in perturbative QFT.
The main point we would like to make is that 
the functional integral is not ill-defined because of the
oscillatory character of the integrand,
but because of the presence of infinitely many
degrees of freedom.
This issue is exactly the same in the Euclideanized theory and in the Lorentzian theory treated by the steepest descent method.

Instead of giving a general proof, we will show by explicit
calculations that this is true on static spacetimes, 
where the notion of Euclidean continuation is unambiguous.
In all cases, the spatial section is a $d$-dimensional
manifold $\Sigma$
with suitable boundary conditions that allow integrations
by parts to be performed without boundary remnants.
We give general formulas for the dependence of the EA on the metric
in $\Sigma$, in any dimension, 
but the main point is already clear when $\Sigma$ 
is a flat torus of side $L$,
in which case the dependence on the metric reduces simply
to the dependence on the total volume $V=L^{d-1}$.

We begin in Section 2, by defining directly the 
Lorentzian functional integral by the steepest descent method.
We then consider in Section 3 the EA of a massive theory 
where time is non-compact.
We prove that the results of the Euclidean and Lorentzian
calculations are indeed related as in Equation (\ref{WEA}).
In Section 4 we consider the case where time is periodic
with period $T$, which is somewhat pathological in the Lorentzian case,
but is useful to illustrate the behavior of massless fields.
Again, we prove that results of the Euclidean and Lorentzian
calculations are related as in Equation (\ref{WEA}),
both in the massive and massless case.
We also discuss the limit $T\to\infty$.
In Section 5 we show how the EA can be calculated by integrating
a Renormalization Group (RG) equation.
This is useful for applications of the RG to gravity.
As a side result we show that the EA calculated in this way
is independent of the choice of the cutoff that enters the definition
of the coarse-graining.
Section 6 contains some final remarks.

\section{Lorentzian functional integral}

Consider a massive scalar field $\phi$ on a $d$-dimensional static spacetime $M$, with action
\begin{equation*}
S[\phi;g] = -\frac{1}{2}\int d^d x\sqrt{-g} 
\left[g^{\mu \nu}\partial_{\mu}\phi\partial_{\nu}\phi
+m^2\phi^2 \right]
=\frac{1}{2}\int d^dx\sqrt{-g}\,\phi(\Box-m^2)\phi \ .
\end{equation*}
Here $\Box=\nabla^2$ is the covariant d'Alembertian, whose 
eigenfunctions will be denoted $\phi_n$,
where $n$ is a composite index comprising an energy eigenvalue
for the Fourier modes in time and another set of labels for
the eigenvalue of the spatial Laplacian $\Delta_\Sigma$.
They are orthonormal 
with respect to the inner product on $C^{\infty}(M)$:
\begin{equation*}
(\phi_n, \phi_m) = \mu^d \int_{M} d^dx\sqrt{-g}\,
\phi_n(x)\phi_m(x) = \delta_{nm}
\end{equation*}
where $\mu$ is a constant with dimension of mass. 
Note that this has nothing to do with the Klein-Gordon
inner product that is used in canonical quantization,
which is only defined for the solutions of the classical
equations of motion.
The eigenfunctions form a basis in the space of functions on $M$, so we can decompose the field $\phi$ as: $\phi(x) = \mu^{(d-2)/2}\sum_n a_n \phi_{n}$. (Note that the coefficients $a_n$ are dimensionless.)
Using the eigenvalue equation
\be
\label{eveq}
\left(\Box-m^2 \right)\phi_n = \lambda_n \phi_n
\ee
the action becomes:
\begin{equation*}
S[\phi;g]=\frac{1}{2\mu^2}
\sum_{n \in \sigma} \lambda_n\, a_n^2 
\end{equation*}
where $\sigma$ is the set that labels the eigenvalues.
The path integral measure can be written formally:
\be
(d\phi) = N \prod_{n \in \sigma} da_n\ ,
\ee
where $N$ is an infinite, field-independent, dimensionless normalization factor, which we define by the 
following Gaussian normalization condition:
\be
1=\int (d\phi)\exp\left[i\frac{\mu^2}{2}
\int d^dx\sqrt{-g}\,\phi^2\right]\ .
\label{measure}
\ee
Using the spectral decomposition of $\phi$,
the r.h.s. becomes a product of integrals of the form
$$
N\prod_{n\in\sigma}
\int_{-\infty}^\infty da_n\exp\left(\frac{i}{2}a_n^2\right)\ .
$$
The integrals over the $a_n$ on the real axis 
are not well-defined, but one can deform the integration
contour in the complex plane to follow the steepest descent
path at the origin, which is shown in Figure 1.
The two eighths of a circle that are needed for this deformation
give no contribution when their radius tends to infinity,
so that each integral is equal to $e^{i\pi/4} \sqrt{2\pi}$. 
Thus we get
\be
N=\prod_{n\in\sigma}\frac{e^{-i\pi/4}}{\sqrt{2\pi}}\ .
\label{exmeasure}
\ee
With the measure (\ref{measure}) and the eigenvalues (\ref{eveq}), 
the partition function becomes a product of integrals
\be
Z_L = N \prod_{n \in \sigma} \int da_n \; 
\exp\left[i\frac{\lambda_n}{2\mu ^2} a_n^2\right]\ .
\ee
Let us split the spectrum into 
$\sigma = \sigma_- \cup \sigma_0 \cup \sigma_+$,
where $\sigma_-$, $\sigma_0$, $\sigma_+$ 
corresponds to $\lambda_n<0$, $\lambda_n=0$, $\lambda_n>0$,
respectively.
In general there will be no zero eigenvalues.
If there is any, the mass has to be given a small imaginary part
$m^2 \to m^2-i\epsilon$, which implies that 
$\lambda_n \to \lambda_n+i\epsilon$: in this way the contribution of the zero eigenvalues becomes a product of standard Gaussian integrals.

\begin{figure}
\begin{center}
\includegraphics[scale=0.5]{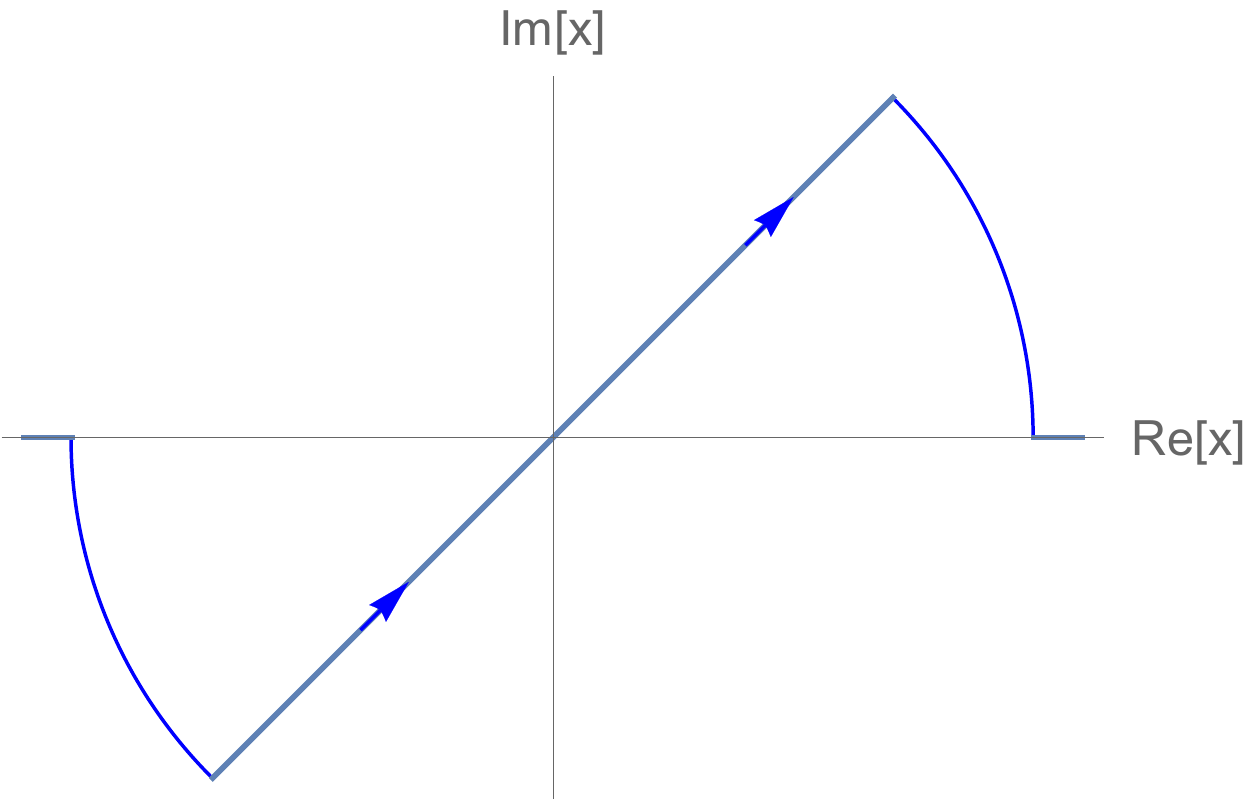}
\quad\quad
\includegraphics[scale=0.5]{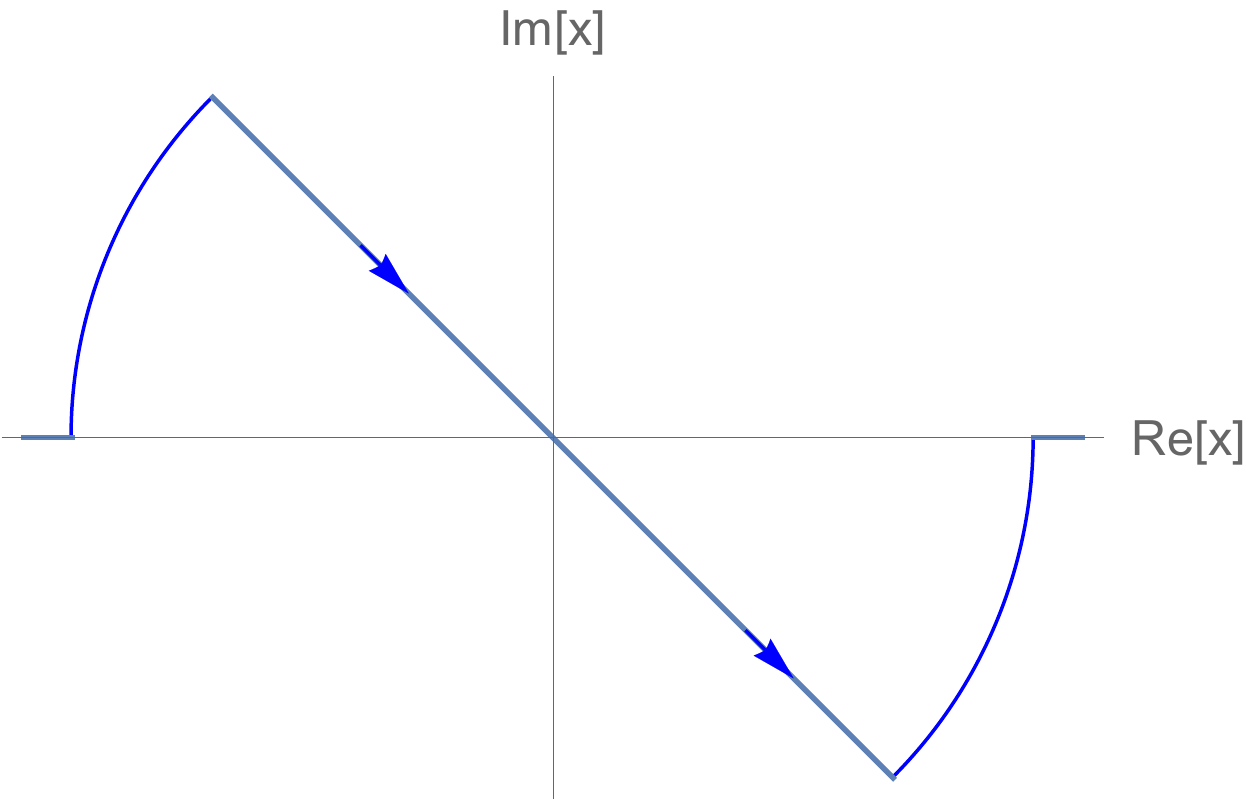}
\caption{Steepest descent paths for the integrals
$e^{i\lambda x^2}$ with $\lambda>0$ (left) and $\lambda<0$ (right).}
\label{fig:1}
\end{center}
\end{figure}

Now let us choose for each integral the path of steepest descent,
shown in Figure 1.
For each positive eigenvalue we get a factor $e^{i\pi/4}$
and for each negative eigenvalue a factor $e^{-i\pi/4}$.
Altogether, using the steepest descent method
and equation (\ref{exmeasure}), the functional integral becomes
\begin{equation}
\begin{split}
Z_L&=N\prod_{n\in \sigma_0} \sqrt{\frac{2\pi \mu^2}{\epsilon}} \prod_{n\in \sigma_-} e^{-i\pi/4} \sqrt{\frac{2\pi \mu^2}{-\lambda_n}}  
\prod_{n \in \sigma_+} e^{i\pi/4}  \sqrt{\frac{2 \pi \mu^2}{\lambda_n}}
\\
&= \prod_{n\in \sigma_0} e^{-i\pi/4} \sqrt{\frac{\mu^2}{\epsilon}} \prod_{n\in \sigma_-} e^{-i\pi/2} \sqrt{\frac{\mu^2}{-\lambda_n}}  \prod_{n \in \sigma_+} \sqrt{\frac{\mu^2}{\lambda_n}}
\\
&=\prod_{n\in \sigma_0} \sqrt{\frac{\mu^2}{i\epsilon}}
\prod_{n\in \sigma_-} \sqrt{\frac{\mu^2}{\lambda_n}}  \prod_{n \in \sigma_+} \sqrt{\frac{\mu^2}{\lambda_n}}
\\
&=\left[ \mbox{det}\left(\frac{\Box-m^2+i\epsilon}{\mu^2}\right) \right]^{-1/2}  \ .
\end{split}
\end{equation}
In the absence of zero eigenvalues, which is the normal situation,
we can drop the $i\epsilon$ and we are led to the standard formula
\be
\Gamma_L=\frac{i}{2}\Tr\log\left(\frac{\Box-m^2}{\mu^2}\right)\ .
\label{gammaL}
\ee
We note that the previous calculation was mathematically
well-defined.
What is ill-defined is only the trace in this last expression,
which is a sum over infinitely many, growing, terms.
This requires regularization and renormalization.
We will discuss this in the next sections, showing that
the results of the direct Lorentzian calculation
agree with the analytic continuation of the results
of the Euclidean calculation.

\section{Non-compact time}

In this section we discuss the case when spacetime has topology
$\mathbb{R}\times \Sigma$ where $\Sigma$ 
is an compact $(d-1)$-dimensional manifold. 
The line element is $ds^2=\pm dt^2+q_{ij}dx^i dx^j$.
The time dependence of the field can be expanded in Fourier integrals
with normal modes $e^{iEt}$ where $-\infty<E<\infty$
is a continuous index.
The trace of the heat kernel of the one-dimensional Laplacian
$-\partial_t^2$ is
\be
\Tr K_T(s)=\frac{1}{2\pi}\int dt\int dE e^{-s E^2}
=\frac{T}{\sqrt{4\pi s}}\ .
\label{trktime}
\ee
where $T=\int dt$ is (infrared) divergent.
This is unavoidable for a static physical system
existing for an infinite time. 
It could be regulated by putting the system in a large ``time box'' 
of finite duration $T$.
In this section we will implicitly assume that this has been done.
In the next section we will discuss the case
when time is periodic with a finite period $T$.

The eigenvalues of the spatial Laplacian 
$\Delta_\Sigma=-q^{ij}\nabla_i\nabla_j$
will be denoted $\omega^2_\alpha$,
where $\alpha$ is a discrete label,
and the trace of its heat kernel has the small-$s$ expansion
\be
\Tr K_\Sigma(s)
=\frac{1}{(4\pi s)^{(d-1)/2}}
\left(B_0(\Delta_\Sigma)+
s B_2(\Delta_\Sigma)
+s^2 B_4(\Delta_\Sigma)+\ldots\right)\ .
\label{trkspace}
\ee
The first expansion coefficients are
\bea
    && B_0(\Delta_\Sigma) = V\\
    && B_2(\Delta_\Sigma) = \frac{1}{6}\int_{\Sigma} d^{d-1}x\;\sqrt{q}\; 
    R\\
    && B_4(\Delta_\Sigma) = \frac{1}{180}\int_{\Sigma} d^{d-1}x\;\sqrt{q}  \left(R_{\mu \nu \rho \sigma}R^{\mu \nu \rho \sigma} -R_{\mu \nu}R^{\mu \nu}+\frac{5}{2}R^2
    -6 \Delta_\Sigma R\right)
\eea
where $V$ is the volume of $\Sigma$ and the curvatures
are those of the metric $q_{ij}$.

\subsection{Euclidean, non-compact time}

The Euclidean EA can be expressed as
\be
\Gamma_E=-\frac{1}{2}\int_0^\infty\frac{ds}{s}\Tr K_{-\Box+m^2}(s)\ ,
\label{master}
\ee
where $K_{-\Box+m^2}(s)=\exp\left[-s(-\Box+m^2)\right]$ is the heat kernel of the (positive) Euclidean kinetic operator
$-\Box+m^2=-\partial_t^2+\Delta_\Sigma+m^2$,
whose eigenvalues are
are
\be
\lambda_{E\alpha} = E^2 + \omega_{\alpha}^2 +m^2\ .
\ee
Since the time and space parts of the Laplacian commute, 
the trace of the heat kernel factors into
\be
\Tr K_{-\Box+m^2}(s)=e^{-m^2 s}\Tr K_T(s)\Tr K_{\Delta_\Sigma}(s)\ .
\ee
Inserting \eqref{trktime} and \eqref{trkspace} in \eqref{master}
we obtain
\bea
\Gamma_{E} 
&=& -\frac{1}{2} \frac{T}{\left( 4\pi \right)^{d/2}} \sum_{j=0}^{\infty} B_{2j}(\Delta_\Sigma)\; \int_{0}^{\infty} ds\,
e^{-m^2 s}   s^{j-\frac{d}{2}-1}
\nonumber\\
&=& -\frac{1}{2}\frac{T m^d}{\left(4\pi\right)^{d/2}} 
\sum_{j=0}^{\infty} \,
\Gamma\left( j-\frac{d}{2}\right)m^{-2j}\,B_{2j}(\Delta_\Sigma) \ .
\label{eancE}
\eea
The integral in the first line is convergent and
gives the result in the second line when $j>d/2$. 
Elsewhere, it can be defined by analytic continuation.
It has poles for $j=0,2,\ldots d$ that can be isolated by
sending $d \to d-\epsilon$.
For example, for $d=4$ \cite{birrelldavies}, 
\bea
\Gamma_E\!\!\!&=&\!\!\!\frac{T}{\left(4\pi\right)^2} 
\left\{ \left(-\frac{1}{\epsilon}+\frac{\gamma}{2}+\log\left(\frac{m}{\sqrt{4\pi}\mu}\right)\right)
\left(\frac{m^4}{2}B_0(\Delta_\Sigma)
-m^2 B_2(\Delta_\Sigma)
+B_4(\Delta_\Sigma)\right)\right.
\nonumber\\
&&
\left.\hspace{2cm}-
\frac{3m^4}{8}B_0(\Delta_\Sigma)
+\frac{m^2}{2} B_2(\Delta_\Sigma)
+\ldots
\right\} \,\ .
\label{dimpoles}
\eea
Alternatively one can put an UV cutoff on the theory by 
integrating $s$ from $1/\Lambda^2$ to infinity.
This gives an incomplete Gamma function:
\be
\Gamma_E=
-\frac{1}{2}\frac{T m^d}{\left(4\pi\right)^{d/2}} 
\sum_{j=0}^{\infty}\,
 \Gamma\left( j-\frac{d}{2},\frac{m^2}{\Lambda^2}\right)m^{-2j} B_{2j}(\Delta_\Sigma)\,\ .
\ee
In the case $d=4$ the expansion of the incomplete Gamma function
gives
\bea
\Gamma_E&=&
-\frac{1}{2}\frac{T}{\left(4\pi\right)^2} 
\Biggl\{
\left[\frac{\Lambda^4}{2}-\Lambda^2 m^2
+\frac{m^4}{4}
\left(3-2\gamma
+4\log\frac{\Lambda}{m}\right)\right]B_0(\Delta_\Sigma)
\nonumber\\
&&
\!\!\!\!\!\!\!\!\!
+\left[\Lambda^2
+m^2\left(-1+\gamma
-2\log\frac{\Lambda}{m}\right)\right]B_2(\Delta_\Sigma)
+\left[-\gamma+2\log\frac{\Lambda}{m}\right]B_4(\Delta_\Sigma)
+\ldots
\Biggr\}\ .
\nonumber
\eea
From here we see that the leading divergences are
powers and logs, multiplied by the heat kernel coefficients,
and we also see that the log divergences exactly reproduce
the dimensional poles in \eqref{dimpoles}.

\subsection{Lorentzian, non-compact time}

In the Lorentzian case the eigenvalues of the kinetic operator
$\Box-m^2=-\partial_t^2-\Delta_\Sigma-m^2$ are
\be
\lambda_{n\alpha} = E^2-\omega_{\alpha}^2-m^2+i\epsilon\ ,
\ee
where we have again added the $i\epsilon$ term in case
that there are zero modes.
The Lorentzian EA is given by the general formula \eqref{master},
where we replace the Euclidean kinetic operator by the
Lorentzian one, and $s$ by $is$:
\be
\Gamma_L=-\frac{i}{2}\int_0^\infty\frac{dis}{is}
\Tr\exp\left[is(\Box-m^2)\right]\ ,
\label{masterL}
\ee
This formula can also be obtained from Schwinger's
action principle, see {\it e.g.} \cite{parkertoms}.
From here we obtain
\footnote{Note that because of the different sign of the temporal part in the eigenvalues the temporal contribution to the Heat Kernel is simply $(-4\pi i s )^{-1/2}$ , while the spatial contribution is 
$(4\pi i s )^{-(d-1)/2}\sum_{j=0}^\infty B_{2j}(\Delta_\Sigma)\,(is)^j$ .}
\bea
\Gamma_L 
&=& -\frac{i}{2} \frac{T}{\left(4\pi\right)^{d/2}} 
\sum_{j=0}^\infty B_{2j}(\Delta_\Sigma)\,
i^{j-\frac{d}{2}+1}
\int_{0}^{\infty} ds\,
e^{-(im^2+\epsilon)s}s^{j-\frac{d}{2}-1}
\nonumber\\
&=& \frac{1}{2}\frac{T m^d}{\left(4\pi\right)^{d/2}} 
\sum_{j=0}^\infty \Gamma\left( j-\frac{d}{2}\right) m^{-2j}\,B_{2j}(\Delta_\Sigma)\,
 \ .
\label{eancL}
\eea
This expression has the same divergences that we encountered
in the Euclidean calculation, and can be regularized
and renormalized in the same way.

For our purposes, we can already compare the Euclidean and
the Lorentzian result.
Recalling that in the standard definition of Wick rotation
$t\to -i t$, we clearly must have $T\to -iT$.
Indeed we see that the two effective actions are related
as in \eqref{WEA}, namely
\be
i\Gamma_L(-iT)=-\Gamma_E(T)\ .
\label{mWEA}
\ee
Thus the direct Lorentzian calculation correctly reproduces 
the analytic continuation of the Euclidean calculation.

\subsection{Zeta function regularization}

In this section we investigate further
the nature of the relation \eqref{mWEA}.
We recall that the determinant of an operator can be calculated using a generalized zeta function \cite{Hawking:1976ja,Elizalde:2007du,Elizalde:2012zza}.
In the non-compact time Euclidean case, it is given by:
\bea
\zeta \left[ z ;  \frac{-\Box_E+m^2}{\mu^2} \right] &=&
\mu^{2z} \sum_\alpha \lambda_\alpha^{-s} =
\frac{\mu^{2z}}{\Gamma(z)}\int_0^\infty ds\; s^{z-1} \Tr K_{-\Box_E+m^2}(s)
\nonumber\\
&=&
\frac{\mu^{2z}}{\Gamma(z)}\frac{T}{(4 \pi)^{d/2}}\sum_{j=0}^\infty \, B_{2j}(\Delta_\Sigma) \int_0^\infty ds\; s^{z+j-\frac{d}{2}-1} e^{-m^2 s}
\nonumber\\
&=&
\frac{\mu^{2z}}{\Gamma(z)}\frac{T m^{d-2z}}{(4 \pi)^{d/2}}
\sum_{j=0}^\infty  \Gamma\left( z+j-\frac{d}{2}\right) m^{-2j}\, B_{2j}(\Delta_\Sigma)
\eea
Then the recipe for the determinant is \cite{Bytsenko:2003tu}:
\be
\log \det \left[ \frac{-\Box_E+m^2}{\mu^2} \right] =
-\lim_{\epsilon\to0}\frac{1}{\epsilon} \zeta \left[ \epsilon ;  \frac{-\Box_E+m^2}{\mu^2} \right] \ .
\nonumber
\ee
Putting $d=4$:
\bea
\log \det \left[ \frac{-\Box_E+m^2}{\mu^2} \right]\!\!\! 
&=&\!\!\!
\frac{T}{\left(4\pi\right)^2} 
\left\{ \left(-\frac{1}{\epsilon}+2\log\frac{m}{\mu}\right)
\left(\frac{m^4}{2}B_0(\Delta_\Sigma)
-m^2 B_2(\Delta_\Sigma)
+B_4(\Delta_\Sigma)\right) \right.
\nonumber\\
&&
\left.\hspace{2cm}-
\frac{3m^4}{4}B_0(\Delta_\Sigma)
+m^2 B_2(\Delta_\Sigma)+\ldots
\right\} \,\ .
\nonumber
\eea
We see that the divergences exactly reproduce
the dimensional poles in \eqref{dimpoles}.

Similarly, in the Lorentzian case: 
\bea
\zeta \left[ z ;  \frac{\Box_L-m^2+i\epsilon}{\mu^2} \right]
\!\!\! &=&\!\!\!
\mu^{2z} \sum_\alpha \left(\lambda_\alpha+i\epsilon \right)^{-s} 
\nonumber\\
&=&\!\!\!
i^{-z}\frac{\mu^{2z}}{\Gamma(z)}\int_0^\infty ds\; s^{z-1} \Tr \exp\left[is(\Box-m^2+i\epsilon)\right]
\nonumber\\
&=&\!\!\!
ie^{-i\pi z}\frac{\mu^{2z}}{\Gamma(z)}\frac{ T m^{d-2z}}{(4 \pi)^{d/2}}
\sum_{j=0}^\infty  \Gamma\left( z+j-\frac{d}{2}\right) m^{-2j}\, B_{2j}(\Delta_\Sigma)\ .
\nonumber
\eea
We see that:
\be
\zeta \left[ z ;  \frac{\Box_L-m^2+i\epsilon}{\mu^2} \right]\Bigg|_{T\to-iT}=e^{-i\pi z}\zeta \left[ z ; \frac{-\Box_E+m^2}{\mu^2}  \right]\ . 
\ee

The formula \eqref{mWEA} can now be seen as a consequence of this relation, since the effective action are proportional to the derivative evaluated for $z=0$ of the zeta function associated to the Hessian of the action.

\section{Compact time}

In this section we discuss a scalar field
in the case when spacetime has topology
$S^1\times \Sigma$ where $S^1$ is a circle of period $T$
and $\Sigma$ is, as before, an compact $(d-1)$-dimensional manifold. 
In this case we will consider both the massive and the massless case.
The time dependence of the field can be expanded in Fourier series
with normal modes $e^{iE_nt}$ where 
$$
E_n=\frac{2\pi n}{T}
$$
and $n\in \mathbb{Z}$.
The eigenvalues of the one-dimensional Laplacian
$-\partial_t^2$ are $E_n^2$ and 
the trace of the heat kernel is
\be
\Tr K_T(s)=\sum_{n=-\infty}^\infty e^{-s E_n^2}\ .
\ee
Note that $T=\int dt$ is finite here, in fact this calculation can be seen
as an infrared regularization of the one performed in
the previous section.
We shall discuss the limit $T\to\infty$ in section 4.4. 

The usual physical interpretation of the Euclidean
EA in $d$ dimensions with a periodic coordinate
is as thermal partition function of a system in
$d-1$ dimensions at temperature $1/T$.
Here, however, we shall also consider the partition function
of a real system in $d$ dimensions with periodic time.
This example is unphysical, but it serves as an illustration
of the fact that partition functions of massless fields
can be calculated directly in Lorentzian signature.

\subsection{Euclidean compact time}

The spectrum of the kinetic operator $-\Box$
for a massive scalar is
\be
\lambda_{n\alpha} = E_n^2 + \omega_{\alpha}^2 + m^2
\label{specE}
\ee
where $n \in \mathbb{Z}$ and $\omega_{\alpha}$ are the eigenvalues of the Laplace-Beltrami operator on $\Sigma$.
The quantum action is given by \eqref{master}.
The heat kernel can now be written
\be
\Tr K_{-\Box+m^2}(s)=\sum_{n,\alpha} e^{-\lambda_{n\alpha}s} =
e^{-m^2 s}\,
\sum_{n=-\infty}^{\infty} e^{-E_n^2s} \,
\Tr K_{\Delta_\Sigma}(s)\ .
\ee

Now using \eqref{master}
and separating the mode $n=0$ from the others
\bea
\Gamma_E &=&
- \frac{1}{2\left( 4\pi \right)^{(d-1)/2}}
\sum_{j=0}^{\infty} B_{2j}(\Delta_\Sigma) \sum_{n=-\infty}^{\infty} 
\int_{0}^{\infty} ds\; 
e^{-\left(E_n^2+m^2\right) s} 
s^{j-\frac{d-1}{2}-1} 
\nonumber\\
&=& 
-\frac{1}{2\left( 4\pi \right)^{(d-1)/2}}
\sum_{j=0}^{\infty} B_{2j}(\Delta_\Sigma)
\sum_{n=-\infty}^{\infty}\left(E_n^2+m^2\right)^{\frac{d-1}{2}-j}\,
\Gamma\left(j-\frac{d-1}{2}\right)
\nonumber\\
&=& 
-\frac{1}{2\left( 4\pi \right)^{(d-1)/2}}
\sum_{j=0}^{\infty} B_{2j}(\Delta_\Sigma)
\left[ m^{d-1-2j}\Gamma\left(j-\frac{d-1}{2}\right)\right.
\nonumber\\
&&\left. +2\left(\frac{2\pi}{T}\right)^{d-1-2j}\sum_{n=0}^{\infty}\frac{(-)^n}{n !}\left(\frac{T m}{2\pi}\right)^{2n}\Gamma\left( j-\frac{d-1}{2}+n\right)\zeta_R\left[2j-d+1+2n\right] \right]\,
\nonumber
\eea
We can thus write
$$
\Gamma_E = \Gamma_E^{T-indep} + \Gamma_E^{T-dep}
$$
where, comparing with \eqref{eancE},
\be
\Gamma_E^{T-indep} = -\frac{1}{2}\frac{m^{d-1}}{\left(4\pi\right)^{(d-1)/2}} 
\sum_{j=0}^\infty \Gamma\left( j-\frac{d-1}{2}\right) m^{-2j}\,B_{2j}(\Delta_\Sigma)
\ee
is equal to the Euclidean effective action of a massive scalar field in the $d-1$ dimensional manifold $\Sigma$ and
\bea
\Gamma_E^{T-dep}\!\!\!\! &=&\!\!\!-\left(\frac{\sqrt{\pi}}{T} \right)^{d-1}\sum_{j=0}^{\infty}\Gamma\left( j-\frac{d-1}{2}\right)\zeta_R\left[2j-d+1\right] 
\left(\frac{T }{2\pi}\right)^{2j}
\sum_{n=0}^{2j}
B_{2j-2n}(\Delta_\Sigma)
\frac{(-)^n m^{2n}}{n !}
\nonumber\\
 &=&\!\!
-\left(\frac{\sqrt{\pi}}{T} \right)^{d-1}\sum_{j=0}^{\infty}\Gamma\left( j-\frac{d-1}{2}\right)\zeta_R\left[2j-d+1\right] 
\left(\frac{T }{2\pi}\right)^{2j}
B_{2j}(\Delta_\Sigma+m^2)
\label{eacdE}
\eea
If we use dimensional regularization to isolate the poles
in this expression, we find that for a given dimension $d$ 
there is only one divergent term in the sum: 
in particular for even $d$ there is a dimensional pole 
when $j=\frac{d}{2}$, coming from the zeta functions,
while for odd $d$ there is a pole for $j=\frac{d-1}{2}$ 
coming from the gamma function.

For example in $d=4-\epsilon$ dimensions
\footnote{Note that in order to have $\Gamma_E$ dimensionless it is necessary 
to $\Gamma_E \to \mu^\epsilon \;\Gamma_E $}
we obtain:
\bea
\Gamma_E &=& -V \left( \frac{\pi^2 }{90 T^3}+\frac{m^3}{48\pi}\right)
-\left( \frac{1}{24T}-\frac{m}{8\pi} \right) B_2(\Delta_\Sigma+m^2)-
\nonumber\\
&& -\frac{T}{16\pi^2}
\left( \frac{1}{\epsilon} + \gamma+\frac{\psi\left(1/2 \right)}{2}
+\ln \frac{T \mu }{\sqrt\pi} \right)
B_{4}(\Delta_\Sigma+m^2)  
+O\left(T^3\right)
\label{4dE}
\eea
In the MS scheme, the finite part of the EA is obtained
by simply dropping the $1/\epsilon$ term.
The first term correctly reproduces the free energy
of a relativistic gas in a box of volume $V$
at temperature $1/T$.
Note that in the case when $\Sigma$ is flat,
there are still infinitely many contributions
coming from the higher $B_{2j}$ that give the
dependence on the mass.

\subsection{Lorentzian compact time}

The spectrum of $\Box$ is
\be
\lambda_{n\alpha} = E_n^2-\omega_{\alpha}^2-m^2+i \epsilon
\label{specL}
\ee
where $n \in \mathbb{Z}$.
Using this in \eqref{masterL} and
proceeding as before and keeping track of the additional
factors of $i$, we find
\bea
\Gamma_L &=& 
\frac{1}{2\left( 4\pi \right)^{(d-1)/2}}
\sum_{j=0}^{\infty} B_{2j}(\Delta_\Sigma)
\left[-i\; m^{d-1-2j}\Gamma\left(j-\frac{d-1}{2}\right)+\right.
\nonumber\\
&&
\!\!\!\!\!\!
\left. +2\;i^{-d}(-1)^j\left(\frac{2\pi}{T}\right)^{d-1-2j}\sum_{n=0}^{\infty}\frac{1}{n !}\left(\frac{T m}{2\pi}\right)^{2n}\Gamma\left( j-\frac{d-1}{2}+n\right)\zeta_R\left[2j-d+1+2n\right] \right]\,
\nonumber
\eea
As before, we can thus write
$$
\Gamma_L = \Gamma_L^{T-indep} + \Gamma_L^{T-dep}
$$
where, comparing with \eqref{eancE},
\be
\Gamma_L^{T-indep} = -\frac{i}{2}\frac{m^{d-1}}{\left(4\pi\right)^{(d-1)/2}} 
\sum_{j=0}^\infty \Gamma\left( j-\frac{d-1}{2}\right) m^{-2j}\,B_{2j}(\Delta_\Sigma)
\ee
is equal to $i$ times the Euclidean effective action of a massive scalar field in the $d-1$ dimensional manifold $\Sigma$ and
\be
\Gamma_L^{T-dep}
= i^{-d}\left(\frac{\sqrt{\pi}}{T} \right)^{d-1}
\sum_{j=0}^{\infty}(-1)^j \,
\zeta_R\left[2j-d+1\right] 
\Gamma\left(j-\frac{d-1}{2}\right)
\left(\frac{T}{2\pi} \right)^{2j}
B_{2j}(\Delta_\Sigma+m^2)
\label{eacdL}
\ee

For example in $d=4-\epsilon$ dimensions we obtain:
\bea
\Gamma_L &=& V \left( \frac{\pi^2 }{90 T^3}-i\frac{m^3}{48\pi}\right)
-\left( \frac{1}{24T}-i\frac{m}{8\pi} \right) B_2(\Delta_\Sigma+m^2)+
\nonumber\\
&& +\frac{T}{16\pi^2}
\left( \frac{1}{\epsilon} + \gamma+\frac{\psi\left(1/2 \right)}{2}
+\ln \frac{i T\mu }{\sqrt\pi} \right)
B_{4}(\Delta_\Sigma+m^2)  
+O\left(T^3\right)
\label{4dL}
\eea
Comparing \eqref{eacdL} with \eqref{eacdE} we see that, in any dimension,
also in this case the EA's are related as in \eqref{WEA},
provided we make the replacement $T\to -iT$.
Note that not only the $T$-independent part but also the $T$-dependent part 
of $\Gamma_L$ is complex 
except for $d$ even, $m=0$ and $\Sigma$ flat .

\subsection{Even vs. odd dimensions}

The overall coefficient in \eqref{eacdL} is real or imaginary,
depending on the dimension. 
For example, for a massless scalar field on a flat torus,
\begin{equation*}
    \Gamma_L = i^{-d} \zeta[1-d] \Gamma \left( \frac{1-d}{2} \right)\,\left(\frac{\sqrt{\pi}}{T} \right)^{d-1}\,V
\end{equation*}
so we see that $\Gamma_L$ is real if $d$ is even 
and imaginary if $d$ is odd.
This is related to the fact that the Euclidean action is real
and contains $T^{1-d}$, so that the Wick rotation
produces a factor $i^d$.
This is in contrast to the case when time is non-compact,
where there is always just one power of $T$ in the EA.

The physical origin of this behavior may be related to the
different propagation properties of fields in even and odd dimensions.
Consider a massless scalar field in 
$\mathbb{R} \times \mathbb{R}^{d-1}$,
satisfying the wave equation 
$$
\left(-\partial_t^2+\vec\partial^2\right) \phi(t,\vec{x})=0\ .
$$
If $d$ is even the value of $\phi$ at a point $(t,\vec{x})$
is determined by its value on the past light cone of the 
point $(t,\vec x)$.
By contrast, if $d$ is odd, then the value of $\phi$ at 
$(t,\vec{x})$
is determined by its value in the interior of the past light cone
\cite{balazs,Dai:2013cwa}.

Now let's compactify the time $\mathbb{R}$ to $S^1$
by identifying $(t,\vec{x})$ and $(t+\frac{2\pi}{T}n,\vec{x})$, where $n \in \mathbb{Z}$.
There are now closed timelike curves, which lead to causality
violation.
However, for a generic point (which means that its coordinates
are irrational multiples of the periods), 
the past light cone never passes through the point itself.
Thus generically there are no closed light rays and therefore
in even dimensions there is no violation of causality for massless fields.
If $d$ is odd the value of the field is affected by its whole
causal past and the presence of closed timelike curves
leads to violation of causality.

\subsection{The limit $T \to\infty$.}

In the infinite volume limit the EA of a massless theory
has infrared divergences.
Therefore, there are two possible ways of taking the limit
$T \to\infty$:
either by separating the mass term from the heat kernel expansion of the spatial part, or by introducing an
explicit IR regulator in the integral over $s$.

Let us first consider the Euclidean case.
Using the identity
\be
\sum_{n=-\infty}^{\infty}  e^{- A n^2} = \sqrt{\frac{\pi}{A}} \left[ 1+ \sum_{n=1}^{\infty}   e^{- \frac{ \pi^2 n^2 }{A}} \right]
\ee
we can write
\bea
\Gamma_E &=&
-\frac{1}{2}\frac{1}{\left( 4\pi \right)^{(d-1)/2}}
\sum_{j=0}^{\infty} B_{2j}(\Delta_\Sigma)
\int_{0}^{\infty} 
ds\, \left[ \sum_{n=-\infty}^{\infty}  e^{-\left( \frac{2\pi n}{T} \right)^2s} \right] e^{-m^2 s} s^{j-\frac{d-1}{2}-1}
\nonumber\\
&=& 
-\frac{1}{2}\frac{1}{\left( 4\pi \right)^{(d-1)/2}}
\sum_{j=0}^{\infty} B_{2j}(\Delta_\Sigma)\int_{0}^{\infty} ds\; \frac{T}{\sqrt{4 \pi s}}\left[ 1+ \sum_{n=1}^{\infty}   e^{-\left( \frac{ n T}{2 \pi \sqrt{s}} \right)^2} \right] e^{-m^2 s} s^{j-\frac{d-1}{2}-1}\ .
\nonumber
\eea
In the limit $T\to\infty$ only the first term 
in the square bracket survives and we arrive at
\be
\Gamma_E =-\frac{1}{2}\frac{T}{\left( 4\pi \right)^{d/2}}\sum_{j=0}^{\infty} B_{2j}(\Delta_\Sigma) 
\int_{0}^{\infty} ds\;e^{-m^2 s} s^{j-\frac{d}{2}-1}\ ,
\ee
which agrees with \eqref{eancE}.
Using the same trick as above, the Lorentzian EA reads
$$
\Gamma_L=-\frac{i}{2} 
\frac{T}{\left( 4\pi \right)^{(d-1)/2}}
\sum_{j=0}^{\infty} i^{j-\frac{d-1}{2}} B_{2j}(\Delta_\Sigma)
\int_{0}^{\infty} ds 
\frac{e^{-im^2 s}}{\sqrt{-4\pi is}}
\left[1+\sum_{n=1}^{\infty}e^{-i\left(\frac{n T}{2\pi\sqrt{s}} \right)^2} \right] 
s^{j-\frac{d-1}{2}-1}\ .
$$
Before taking the limit, we have to remember the $i\epsilon$ prescription $m^2 \to m^2 -i\epsilon$ and $E_n \to E_n +i\epsilon$, which is equivalent to $T \to T -i\epsilon$.
If we keep $\epsilon$ fixed, the second term in the square bracket becomes 
$e^{-i\left(\frac{n T}{2\pi\sqrt{s}} \right)^2} \to e^{-i\left(\frac{n T}{2\pi\sqrt{s}} \right)^2-\left(\frac{n }{2\pi\sqrt{s}} \right)^2 \epsilon T +o(\epsilon^2)}$.
As before, in the limit $T\to\infty$, only the first term 
in the square bracket survives and then taking $\epsilon\to0$ 
we arrive at \eqref{eancL}.

\section{Deriving the EA from an RG equation}

The Effective Average Action (EAA) is the EA for a theory
where the action has been modified by the addition of a 
``cutoff term'' $S_k$ that suppresses the contribution
of low-momentum modes.
It depends on the ``IR cutoff'' $k$ and reduces to the
ordinary EA when $k\to 0$.
The cutoff term has the general form
$$
S_k=\int d^dx\sqrt{g}\phi R_k(z)\phi\ ,
$$
where $z$ is a suitable second order differential operator
and the function $R_k$ tends to $k^2$ for $k\to0$
and goes rapidly to zero for $z>k$.
The ``cutoff kinetic operator'' is then
$$
P_k(z)=z+R_k(z)\ .
$$
The choice of the regulator function $R_k$ is largely arbitrary
and is somewhat analogous to choosing a renormalization scheme in perturbation theory. 
At one loop the EAA is then given by
\be
\Gamma_{k}=\frac{1}{2}\Tr\log\left(\frac{P_k(z)}{\mu^2}\right)\ .
\ee
Note that the one-loop EAA $\Gamma_k$ is ill-defined
and needs UV regularization and renormalization.
However, if we take a derivative with respect to $k$,
the resulting expression is UV finite.
It can be interpreted as a RG equation
\cite{wett1,morris1}:
\be
k\frac{\partial\Gamma_k}{\partial k}=\frac{1}{2}
\Tr\left(P_k(z)^{-1}\, k\frac{\partial R_k(z)}{\partial k}\right)\ .
\label{erge}
\ee
The EA can be calculated by integrating the flow from
some UV scale $k=\Lambda$ to $k=0$. 
For some concrete examples see \cite{Codello:2015oqa}.
The usual divergences of QFT
are hidden in the choice of the initial condition for the EAA
at the UV scale and reappear when we try to send $\Lambda\to\infty$.

Normally, in a Euclidean setting, one chooses $z$ to be
a Laplace-type operator. This preserves rotational invariance
and guarantees that the cutoff represents physically
a coarse-graining of the degrees of freedom of the system.
The RG equation \eqref{erge} is much less used in a Lorentzian setting.
One may try to introduce a cutoff preserving Lorentz invariance by choosing 
$z$ to be a d'Alembertian operator, but in this way there
would be no restriction on the modulus of the spatial momenta.
In other words, one would not be really coarse-graining in the 
usual sense.
Alternatively, one could impose separate cutoffs on the
space and time components of the momentum,
so that only low wave numbers and low frequencies are suppressed.
In this way one would lose Lorentz covariance.
For a general discussion see \cite{Floerchinger:2011sc,Pawlowski:2015mia}.

\subsection{Compact time with optimized regulator}

For the calculations of the EA in periodic time,
where Euclidean/Lorentz invariance is broken anyway by the static metric,
it will be adequate to cut off only the space momenta
\cite{Manrique:2011jc,Rechenberger:2012dt}.
In this section we will treat the Euclidean and Lorentzian
calculations together.
The kinetic operator is $z=z_T+\sigma z_\Sigma$,
where $z_T=-\partial_t^2$, $z_\Sigma=\Delta_\Sigma$
are both positive operators,
and $\sigma=1$ in the Euclidean case
while $\sigma=-1$ in the Lorentzian case.
We choose a cutoff of the form
$\sigma R_k(z_\Sigma)$, so that the cutoff kinetic operator is
$z_T+\sigma z_\Sigma+\sigma R_k(z_\Sigma)=z_T+\sigma P_k(z_\Sigma)$,
where $P_k(z)=z+R_k(z)$.
Thus the RG equation \eqref{erge} becomes
\be
k\frac{\partial\Gamma_k}{\partial k}=\frac{\sqrt\sigma}{2}
\Tr\left(\frac{\sigma}{\sigma P_k(z_\Sigma)+ z_T}\, 
k\frac{\partial R_k(z_\Sigma)}{\partial k}\right)\ ,
\label{erge2}
\ee
where the prefactor $\sqrt\sigma$ accounts for the
$i$ in \eqref{gammaL} in the Lorentzian case.
As we have seen in the preceding examples, the trace
factors as a product of two traces, over the time and space
quantum numbers.
Even though we have not put a cutoff on the time quantum numbers,
both traces are now finite, as we shall see below.

With our choice of cutoff $R_k(\Delta_\Sigma)$
and with the eigenvalues of equation \eqref{specE} and \eqref{specL},
the Euclidean RG equation reads
\be
k\frac{\partial\Gamma_k}{\partial k}=\frac{\sqrt\sigma}{2}
\sum_{n,\alpha}
\left(\frac{\sigma}{\sigma P_k(\omega_\alpha^2)+E_n^2}\, \, 
k\frac{\partial R_k(\omega_\alpha^2)}{\partial k}\right)\ ,
\label{arianna}
\ee
Let us first see what this gives for $\Sigma=\mathbb{R}^{d-1}$.
The sum over $\alpha$ then becomes an integral over space momenta $p$.
This leads to the formula
\be
k\frac{\partial\Gamma_k}{\partial k}
=\frac{\sqrt\sigma}{2}\frac{V}{(4\pi)^{(d-1)/2}}
\sum_{n=-\infty}^\infty
Q_{\frac{d-1}{2}}\left[\frac{\sigma}{\sigma P_k+ E_n^2}\, 
k\frac{\partial R_k}{\partial k}\right]\ ,
\label{carlo}
\ee
where
\be
Q_m[W]=\frac{1}{\Gamma(m)}\int dz z^{m-1}W(z)\ .
\ee
Defining $y=|p|^2/k^2$ and $R_k(p^2)=k^2 r(y)$ where $y=p^2/k^2$,
the relevant $Q$-functionals can be rewritten as
\be
Q_m\left[\frac{\sigma}{\sigma P_k+ E_n^2}\, 
k\frac{\partial R_k}{\partial k}\right]
=\sigma \frac{2k^{2m}}{\Gamma(m)}\int dy y^{m-1}
\frac{r(y)-yr'(y)}{\sigma(y+r(y))+ \tilde E_n^2}\ ,
\ee
where $\tilde E_n=E_n/k$.
In general these functionals depend on the shape of the
function $r(y)$, but in the case $m=0$ 
(corresponding to $d=1$) they do not \cite{cpr2}.

In general, the $Q$-functional can be calculated explicitly
for the optimized cutoff $r(y)=(1-y)\theta(1-y)$:
\be
Q_{\frac{d-1}{2}}\left[\frac{\sigma}{\sigma P_k+ E_n^2}\, 
k\frac{\partial R_k}{\partial k}\right]
=\frac{2k^{d-1}}{\Gamma\left(\frac{d+1}{2}\right)}
\frac{\sigma}{\sigma+ \tilde E_n^2}\ .
\ee
Then using
$$
\sum_{n=-\infty}^\infty \frac{1}{\sigma+\tilde E_n^2}
=\frac{T k}{2\sqrt{\sigma}}\coth\frac{\sqrt{\sigma}T k}{2}\ ,
$$
we arrive at
\be
k\frac{\partial\Gamma_k}{\partial k} =
\frac{\sigma T\, Vk^{d}}{2\Gamma\left(\frac{d+1}{2}\right)(4\pi)^{(d-1)/2}} \coth\frac{\sqrt{\sigma}T k}{2}\ .
\label{elena}
\ee

Let us specialize to the case $d=4$.
Integrating the LHS of (\ref{elena}) we obtain 
$\Gamma_\Lambda-\Gamma_0$, where $\Lambda$ is an arbitrary UV scale:
\begin{eqnarray}
\Gamma_\Lambda-\Gamma_0&=& 
\sigma \frac{V T}{12\pi^2}
\left[ -\frac{2 \pi ^4}{15 T ^4}
-\frac{\Lambda ^4}{4}
+\frac{2\Lambda^3\log\left(1-e^{\sqrt\sigma T\Lambda }\right)}{\sqrt\sigma T} \right.
\nonumber\\
&+& \left.
\frac{12 \mathrm{Li}_4\left(e^{\sqrt\sigma T  \Lambda }\right)}{T ^4}
-\frac{12\sqrt\sigma \Lambda  \mathrm{Li}_3\left(e^{\sqrt\sigma T  \Lambda }\right)}{ T^3}
+	\frac{6\sigma \Lambda ^2 \mathrm{Li}_2\left(e^{\sqrt\sigma T  \Lambda }\right)}{T ^2} \right]\ .
\end{eqnarray}
For $\sigma=1$, expanding the last four terms in the square bracket around
$\Lambda=\infty$ gives
$$
\frac{4\pi^4}{15 T^4}+\frac{\Lambda^4}{2}
+O(e^{-T \Lambda})\ .
$$
In the limit $\Lambda\to\infty$ we thus obtain
\be
\Gamma_\Lambda-\Gamma_0 = 
\frac{V T}{12\pi^2}
\left[\frac{2 \pi ^4}{15 T^4}
+\frac{\Lambda ^4}{4}
+O(e^{-T\Lambda}) \right]\ .
\ee
In order to extract the EA $\Gamma\equiv\Gamma_0$, 
we need to specify the initial condition $\Gamma_\Lambda$.
For $\Lambda\to\infty$ it must reproduce the classical action,
which in the case of vanishing field that we are considering here
is simply zero.
This means that $\Gamma_\Lambda$ must be chosen to be equal
to the divergent term on the r.h.s., and the final result is
\be
\label{eq:EAAe-result-FRG}
\Gamma_{0E}= -\frac{V_3\pi^2}{90 T^3}\ .
\ee
in accordance with (\ref{4dE}).
For $\sigma=-1$, 
we have to remember that there is a $i\epsilon$ prescription $E_n \to E_n +i\epsilon$ which is equivalent to $T \to T -i\epsilon$: as a consequence the exponential factor in the polylogs function becomes $e^{i T  \Lambda } \to e^{(i T   +\epsilon) \Lambda}$. So we have the same expansion as before and we obtain
\be
\Gamma_{0L}=\frac{V_3\pi^2}{90 T^3}\ .
\ee
in accordance with (\ref{4dL}).

\subsection{Compact time with general regulator}

The sums in \eqref{arianna} can be treated also for general $\Sigma$
using the general formula for the trace of a function of a Laplacian,
given in appendix A of \cite{cpr2}.
One obtains
\be
k\frac{\partial\Gamma_k}{\partial k}
=\frac{\sqrt\sigma}{2}\frac{1}{(4\pi)^{(d-1)/2}}
\sum_{n=-\infty}^\infty
\sum_{j=0}^\infty
Q_{\frac{d-1}{2}-j}\left(\frac{\sigma}{\sigma P_k+ E_n^2}\, 
k\frac{\partial R_k}{\partial k}\right)B_{2j}(\Delta_\Sigma)\ ,
\label{genrg}
\ee
whose first term indeed agrees with \eqref{carlo}.

In the preceding calculation we had to choose a special
regulator in order to be able to perform all the integrals
and sums in closed form.
However, it is possible to show that the final result
for the EA is actually independent of the choice of
cutoff function $R_k$.

Since the integral over $s$ and the sum over $n$ cannot
be performed for a general cutoff,
in order to show cutoff-independence
we have to perform the integration over $k$
{\it before} the integral over $y$ and the sum over $n$.
This inversion is legitimate because the sums and
integrals are convergent.

Let us go back to equation \eqref{genrg} and integrate both sides.
In the r.h.s. we encounter the integrals
$$
\int_0^\Lambda \frac{dk}{k}
Q_{\frac{d-1}{2}-j}\left[\frac{\sigma}{\sigma P_k+ E_n^2}
k\frac{\partial R_k}{\partial k}\right]
=\frac{2}{ \Gamma\left(\frac{d-1-2j}{2}\right)}
\int\! dy y^{\frac{d-3}{2}-j}
\frac{r(y)-yr'(y)}{y+r(y)}
\int_0^\Lambda\! dk\frac{k^{d-2j}}{k^2+\frac{ E_n^2}{\sigma(y+r)}}
$$
The $k$-integral can be performed exactly, yielding a hypergeometric function. For large $\Lambda$ it diverges,
leaving a finite part
$$
\frac{2\sigma^{-(d-1-2j)/2}}{d+1-j}\Gamma\left(\frac{1+2j-d}{2}\right)
\Gamma\left(\frac{d+3-2j}{2}\right)
\frac{E_n^{d-1-2j}}{(y+r)^{(d-1-2j)/2}}\ .
$$
With this, the $y$-integral is seen to become the
integral of a total $y$-derivative,
and therefore becomes independent of the shape of
the cutoff function.
The final result is
$$
\Gamma\left(\frac{1+2j-d}{2}\right)
\sigma^{-(d-1-2j)/2}\,
E_n^{d-1-2j}\ .
$$
Inserting this back in \eqref{genrg} and performing
the sum over $n$ one obtains
\be
\Gamma_0=
-\sigma^{1-\frac{d}{2}}\left(\frac{\sqrt{\pi}}{T}\right)^{d-1}
\sum_{j=0}^\infty \sigma^j\,
\zeta_R[1-d+2j]\, \Gamma \left( \frac{1-d}{2}+j \right)
\left(\frac{T}{2\pi}\right)^{2j}
B_{2j}(\Delta_\Sigma)\ .
\ee
which agrees with equations (\ref{eacdE},\ref{eacdL}) for $\sigma=\pm1$ and $m=0$.

\subsection{Non-compact time with general regulator}

To pass from the compact case to the non-compact one we just do the following substitution:
$$
E_n \to \sqrt{E^2 + \sigma m^2} 
\hspace{1cm}\mbox{and}\hspace{1cm}
\sum_n \to \frac{1}{2\pi} \int dt \int_{-\infty}^{\infty} dE = \frac{T}{2\pi} \int_{-\infty}^{\infty} dE
$$
The integral in $E$ gives:
$$
\int_{-\infty}^{\infty} dE \left(E^2+\sigma m^2\right)^{(d-1-2j)/2} = \sigma^{(d-2j)/2} \sqrt{\pi} \frac{\Gamma \left(  j-\frac{d}{2}\right)}{\Gamma \left(\frac{1+2j-d}{2}\right)} m^{d-2j} 
$$
Finally:
\be
\Gamma_0=
-\frac{\sigma}{2}\frac{T m^d}{\left(4\pi\right)^{d/2}} 
\sum_{j=0}^\infty \Gamma\left( j-\frac{d}{2}\right)m^{-2j}\, B_{2j}(\Delta_\Sigma)\,
 \ ,
\ee
which agrees with equations (\ref{eancE},\ref{eancL}) for $\sigma=\pm1$.

\section{Discussion}

We have discussed the definition of functional integrals
directly in Lorentzian signature, by means of the steepest
descent procedure.
We conclude with two remarks.
We have considered here only Gaussian
functional integrals.
For an interacting theory the perturbative expansion
can be derived from the expression
\bea
Z_\sigma[j]&=\exp\left[\sqrt{\sigma}S_\mathrm{int}\left(\frac{1}{\sqrt{\sigma}}\frac{\delta}{\delta j}\right)\right]
\int (d\phi)\exp\left[\sqrt{\sigma}\int d^dx\sqrt{g}\,\left(\frac{1}{2}\phi \Delta_\sigma \phi +j\phi\right)\right]=
\nonumber\\
&=Z_\sigma[0]\exp\left[\sqrt{\sigma}S_\mathrm{int}\left(\frac{1}{\sqrt{\sigma}}\frac{\delta}{\delta j}\right)\right]
\exp\left[-\frac{\sqrt{\sigma}}{2}\,j\cdot\Delta_\sigma^{-1}\cdot j\right]
\eea
where $\Delta_\sigma$ is the Laplacian for $\sigma=1$ and the d'Alembertian for $\sigma=-1$.
The only path integral that one has evaluated is a Gaussian one.
Thus the steepest-descent procedure is enough to define
the Lorentzian theory perturbatively.
It is only when one wants to go beyond the perturbative treatment
that more complicated contours will have to be chosen,
such as in \cite{Feldbrugge:2017kzv,Basar:2013eka}.

In the case of quantum gravity, as noted in the introduction,
there is the issue that the free action of gravitons has
opposite sign for the spin-two and for the spin-zero components. 
We note that the steepest-descent prescription implies different
integration contours for these fields,
each leading to a determinant.
Thus we arrive at the same result as the analytic
continuation of the Euclidean integral,
treated with the Cambridge prescription.
\bigskip

{\bf Acknowledgements}. 
R.P. wishes to thank the organizers of the conference
``The path integral for gravity'', which was held at Perimeter Institute, November 13-17, 2017, and provided the motivation
for this work.
This research was supported in part by Perimeter Institute for Theoretical Physics. Research at Perimeter Institute is supported by the Government of Canada through Industry Canada and by the Province of Ontario through the Ministry of Economic Development and Innovation.


\end{document}